\documentclass[a4 paper, 12 pt] {article}

\usepackage{graphics, graphicx}

\begin{document}
\title{\bf{Nested Multi-soliton Solutions with Arbitrary Hopf Index}}
\author{A. Wereszczy\'{n}ski $^{a)}$ \thanks{wereszcz@alphas.if.uj.edu.pl}
       \\
       \\ $^{a)}$ Institute of Physics,  Jagiellonian University,
       \\ Reymonta 4, Krak\'{o}w, Poland}
\maketitle
\begin{abstract}
Generalized Aratyn-Ferreira-Zimerman $O(3)$ nonlinear sigma
mo\-del with a particular symmetry breaking term, so-called
dielectric function, is discussed. Static multi-soliton
configurations with finite energy and nontrivial Hopf index are
found. We show that such configurations consist of nested toroidal
solitons. Moreover, nontrivial sphaleron-like solutions i.e.
configurations with zero total topological charge are also
presented.
\end{abstract}
\newpage
\section{\bf{ Introduction}}
It is widely believed that toroidal knotted topological solitons
so-called hopfions play a very important role in the temporary
physics. In fact, they seem to give a natural language of
describing particle excitations in the low energy sector of the
quantum gluodynamics i.e. famous glueball states. This idea,
proposed and developed by Faddeev, Niemi \cite{niemi},
\cite{langmann} and Cho \cite{cho}, provides an elegant framework
where masses as well as other physical properties can be
understood in terms of topological characteristics of solutions
\footnote[1]{For other applications of knotted solitons see
\cite{babaev}.}. In particular, the well-known
Vakulenko-Kapitansky inequality \cite{vakulenko} gives the
following spectrum of the glueballs $E \sim 1500 MeV
|Q_H|^{\frac{3}{4}}$, where $Q_H=1,2...$ is Hopf index. Moreover,
it has been also observed that the spectrum should possess
right-left degeneracy \cite{niemi_new}. It follows from the
observation that energy of the topological solutions do
not depend on the sign of the topological charge. \\
Unfortunately, in case of the Faddeev-Niemi model only numerical
solutions have been found \cite{battyde}, \cite{salo},
\cite{ward}. However, there exist models which allow us to learn
something more about the mathematical structure and behavior of
toroidal solutions \cite{nicole}. They can be regarded as a toy
models where we can test some ideas borrowed from the standard
soliton theory in two space dimensions. In fact, the
Aratyn-Ferreira-Zimerman \cite{aratyn} model is a widely discussed
example of a theory where exact soliton solutions with arbitrary
Hopf number have been obtained. Moreover, some generalizations to
$N$-interacting Aratyn-Ferreira-Zimerman-like models \cite{my3} or
including a symmetry breaking term have been discussed and
analytically solutions obtained \cite{my2}. In particular, the
problem of the existence of hopfions in models with broken $O(3)$
global symmetry seems to play the important role in the context of
gluodynamics. It follows from the fact that the Faddeev-Niemi
model admits massless excitations being an effect of the
spontaneous $O(3)$ symmetry breaking. This pathological behavior
can be cured by adding some explicitly symmetry breaking term in
the action, see for example \cite{niemi2}, \cite{sanchez1},
\cite{wipf}, \cite{my1}. In general such modified Faddeev-Niemi
lagrangian has the form
\begin{equation}
\mathcal{L}=-\frac{\sigma_1 (\vec{n} ) }{4} [ \vec{n} \cdot (
\partial_{\mu } \vec{n} \times \partial_{\nu } \vec{n} )]^2 +
 \frac{\sigma_2 (\vec{n})}{2} (\partial_{\mu } \vec{n} )^2, \label{lag_fn2}
\end{equation}
where two so-called dielectric functions $\sigma_1$ and $\sigma_2$
have been introduced. Moreover, one can also include a potential
term for $\vec{n}$. Obviously this modification makes the original
model even more complicated and, so far, no analytical
calculations have been presented.
\\
In the present work we would like to continue the investigation of
hopfions in models with broken global $O(3)$ symmetry. In order to
do it, we will take advantage of the Aratyn-Ferreira-Zimerman
model \cite{aratyn} with a particular symmetry breaking dielectric
function \cite{my2}.
\\
The main aim of our work is to analyze multi-soliton
configurations in this toy model. We are especially interested in
construction of sphaleron-like solutions i.e. configurations with
zero total topological charge but non-trivial local topological
structure. Such solutions might be helpful in finding of a time
depending topological soliton i.e breather, which in general
consists of one soliton and one antisoliton component.
Additionally, it would give us also a chance to investigate the
decay and scattering of hopfions.
\\
Due to the fact that the symmetry breaking is realized in the same
manner as in the QCD motivated model (\ref{lag_fn2}) our work
might be regarded us the first step in analytical investigation of
the scattering and decay of glueballs as well as in finding of the
breather which can change the spectrum of the glueballs.
\section{\bf{ Multi-Soliton Configuration}}
In this paper we will look for toroidal topologically nontrivial
configurations in $(3+1)$ Minkowski space-time for the following
Lagrangian density \cite{my2}, \cite{sanchez2}
\begin{equation}
\mathcal{L}= \sigma \left(\vec{n} \right) \left[ [ \vec{n} \cdot (
\partial_{\mu } \vec{n} \times
\partial_{\nu } \vec{n} )]^2 \right]^{\frac{3}{4}},
\label{toylag}
\end{equation}
where the symmetry breaking dielectric function $\sigma $,
so-called dielectric function, is chosen in a very special form
which provides analytical solutions
\begin{equation}
\sigma (\vec{n})=\frac{1}{(1-(n^3)^2)^{\frac{3}{4}}}.
\label{sigma}
\end{equation}
$\vec{n}=(n^1,n^2,n^3)$ is an unit three component vector field.
As it was shown in \cite{my2} such a model belongs to a wide
family of integrable theories. Here, integrability is understood
in the sense that infinitely many conserved currents exist
\cite{alvarez}, \cite{ferreira}.
\\
In order to find static soliton solutions we take advantage of the
stereographic projection
\begin{equation}
\vec{n}= \frac{1}{1+|u|^2} ( u+u^*, -i(u-u^*), |u|^2-1)
\label{stereograf}
\end{equation}
and introduce toroidal coordinates
$$ x=\frac{a}{q} \sinh \eta \cos \phi , $$
$$ y=\frac{a}{q} \sinh \eta \sin \phi , $$
\begin{equation}
z=\frac{a}{q} \sin \xi ,\label{tor_coord}
\end{equation}
where $q=\cosh \eta -\cos \xi $ and $a>0$ is a constant of
dimension of length fixing the scale in the coordinates. Moreover,
we use Aratyn-Ferreira-Zimerman Ansatz \cite{aratyn}
\begin{equation}
u(\eta,\xi,\phi) \equiv f(\eta) e^{i(m\xi + n \phi)},
\label{anzatz}
\end{equation}
where $m, \; n$ are integers.
\\
Then the static equation of motion reads as follow \cite{my2}
\begin{equation}
\partial_{\eta} \ln \frac{\sigma^{2/3} ff'}{(1+f^2)^2} =-\frac{2m^2 \sinh^2 \eta -n^2}{m^2 \sinh^2 \eta
+n^2} \frac{\cosh \eta}{\sinh \eta}. \label{eqmot-toy4a}
\end{equation}
It can be integrated and we find
\begin{equation}
\frac{\sigma^{2/3} ff'}{(1+f^2)^2} = \frac{k_1}{|m|^3} \frac{\sinh
\eta}{\left( \frac{n^2-m^2}{m^2} +\cosh^2 \eta \right)^{3/2}},
\label{eqmot-toy4}
\end{equation}
where $k_1$ is a constant. In case of the previously introduced
dielectric function this equation can be rewritten as
\begin{equation}
\int \frac{1}{(1+f^2)} df= \frac{-k_1}{|m|(m^2-n^2)} \frac{\cosh
\eta}{\left( \frac{n^2-m^2}{m^2} +\cosh^2 \eta \right)^{1/2}}
-\frac{k_2}{2}, \label{eqmot-toy5}
\end{equation}
and integrated. We obtain the general solution
\begin{equation}
\arctan f =\frac{-k_1}{|m|(m^2-n^2)} \frac{\cosh \eta}{\left(
\frac{n^2-m^2}{m^2} +\cosh^2 \eta \right)^{1/2}} -\frac{k_2}{2},
\label{sol1}
\end{equation}
where $k_2$ is a second integration constant. To fix the value of
the integration constants one has to specify the asymptotic
conditions. They can be chosen as
\begin{equation}
\vec{n} \rightarrow (0,0,-1) \; \; \mbox{i.e.} \; \; f \rightarrow
0 \; \; \mbox{as} \; \; \eta \rightarrow 0 \label{bound1}
\end{equation}
and
\begin{equation}
\vec{n} \rightarrow (0,0,1) \; \; \mbox{i.e.} \; \; f \rightarrow
\infty  \; \; \mbox{as} \; \; \eta \rightarrow \infty.
\label{bound2}
\end{equation}
Then, after some algebra one can find
\begin{equation}
\arctan f =  \frac{\pi}{2} \frac{ (2l+1) }{|m|-|n|} \left( |m| -
|n| \frac{\cosh \eta}{\left( \frac{n^2}{m^2} +\sinh^2 \eta
\right)^{1/2}} \right). \label{sol2}
\end{equation}
In other words we have obtained a whole family of solutions of the
equation (\ref{eqmot-toy5}) which fulfill the assumed asymptotic
conditions. This family is label by the positive and integer
parameter $l=0,1,2...$. Thus, our solutions are given by the
formula
\begin{equation}
f= \tan \left[  \frac{\pi}{2} \frac{ (2l+1)}{|m|-|n|} \left( |m| -
|n| \frac{\cosh \eta}{\left( \frac{n^2}{m^2} +\sinh^2 \eta
\right)^{1/2}} \right)
 \right]. \label{sol3}
\end{equation}
This correspond with the following formula for the $n^3$ component
of the unit field
\begin{equation}
n^3=1 - \frac{2}{1+ \tan^2 \left[  \frac{\pi}{2} \frac{
(2l+1)}{|m|-|n|} \left( |m| - |n| \frac{\cosh \eta}{\left(
\frac{n^2}{m^2} +\sinh^2 \eta \right)^{1/2}} \right)
 \right] }. \label{sol_n3}
\end{equation}
One can see that $n^3$ starts in -1 and tends to +1. Additionally,
it flips $2l+1$ times between -1 and +1. The points, where
$n^3=-1$ define the positions of the solitons. More precisely, the
solution describe a soliton if $n^3$ increases from $-1$ to $+1$.
Analogously, antisoliton appears when $n^3$ decreases from $+1$ to
$-1$. Thus, there are $2l+1$ nested toroidal solitons.
\\
Let us now find the energy corresponding to the solutions. One
obtains
\begin{equation}
E \equiv \int d^3x T_{00} = (2\pi)^2 8 \cdot 2^{3/4}
\int_0^{\infty} \frac{d \eta \sinh \eta}{(1+f^2)^{3}} \left( m^2
+\frac{n^2}{\sinh^2 \eta } \right)^{\frac{3}{4}} f^{\frac{3}{2}}
f'^{\frac{3}{2}} \sigma (f). \label{energy1}
\end{equation}
Inserting our solutions into (\ref{energy1}) we find that the
energy is finite and given by the expression
\begin{equation}
E_{m,n}^l=(2\pi)^2 4 \cdot 2^{1/4} (2l+1)^{3/2}
\sqrt{|m||n|(|m|+|n|)}. \label{energy2}
\end{equation}
The behavior of $n^3$ and distribution of the energy density in
case of $m=2, n=1$ is shown in Fig. 1 and Fig. 2.

\vspace{0.5cm} \centerline{\includegraphics{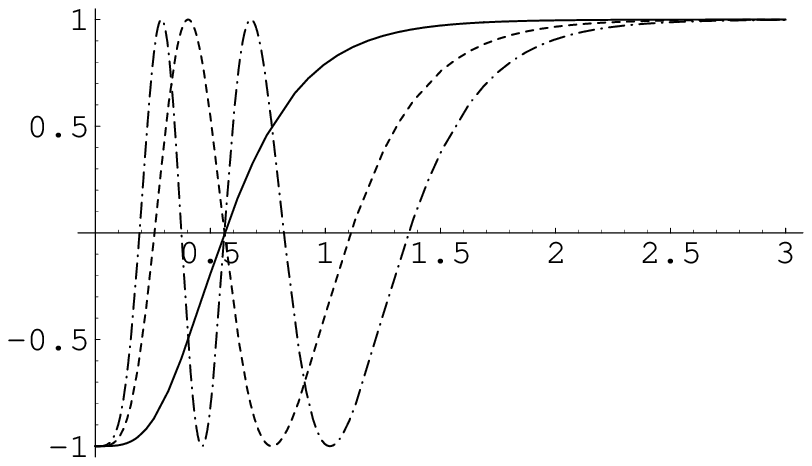}} {\bf
Fig. 1} \small{$ n^3 (\eta ) $ for $l=0,1,2$ - solid, dashed,
dot-dashed line respectively}

\vspace{0.5cm} \centerline{\includegraphics{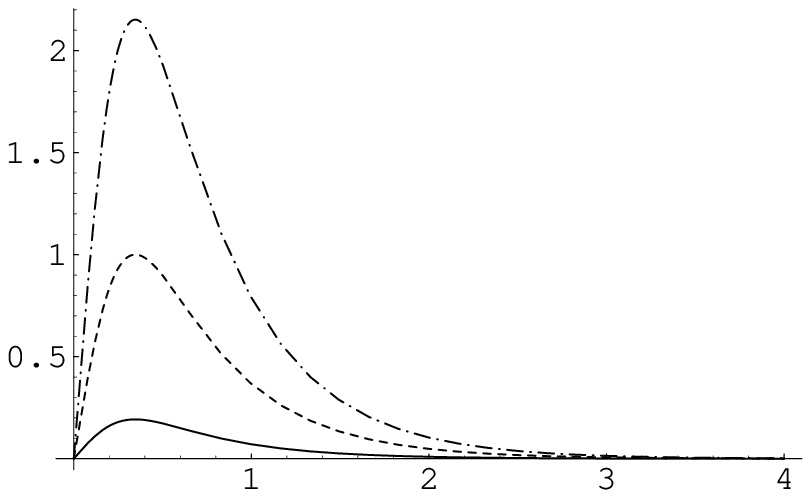}}
{\bf Fig. 2} \small{Energy density for $l=0,1,2$ - solid, dashed,
dot-dashed line respectively} \vspace{0.5cm}

\noindent It is straightforward to see that also the following
asymptotic conditions can lead to multi-soliton configurations
\begin{equation}
\vec{n} \rightarrow (0,0,-1) \; \; \mbox{i.e.} \; \; f \rightarrow
0 \; \; \mbox{as} \; \; \eta \rightarrow 0 \label{bound3}
\end{equation}
and
\begin{equation}
\vec{n} \rightarrow (0,0,-1) \; \; \mbox{i.e.} \; \; f \rightarrow
0 \; \; \mbox{as} \; \; \eta \rightarrow \infty. \label{bound4}
\end{equation}
Here, in contradiction to the case discussed above, the value of
$n^3$ in the center of the torus and in the spatial infinity is
identical. One can check that solutions form a family also label
by positive, integer number $k=0,1,2...$
\begin{equation}
n^3=1 - \frac{2}{1+ \tan^2 \left[   \frac{\pi }{2} \frac{ 2k
}{|m|-|n|} \left( |m| - |n| \frac{\cosh \eta}{\left(
\frac{n^2}{m^2} +\sinh^2 \eta \right)^{1/2}} \right)
 \right] }. \label{sol_n3a}
\end{equation}
Now, $n^3$ flips $2k$ times between $+1$. It is equivalent to the
fact that there are even number of the nested toroidal solitons
and, as we prove it below, the total topological charge vanishes.
Such solutions possess the following total energy
\begin{equation}
E_{m,n}=(2\pi)^2 4 \cdot 2^{1/4} (2k)^{3/2}
\sqrt{|m||n|(|m|+|n|)}. \label{energy3}
\end{equation}
In Fig. 3 and Fig. 4 $n^3$ and energy density for $m=2, n=1$ are
shown.

\vspace{0.5cm} \centerline{\includegraphics{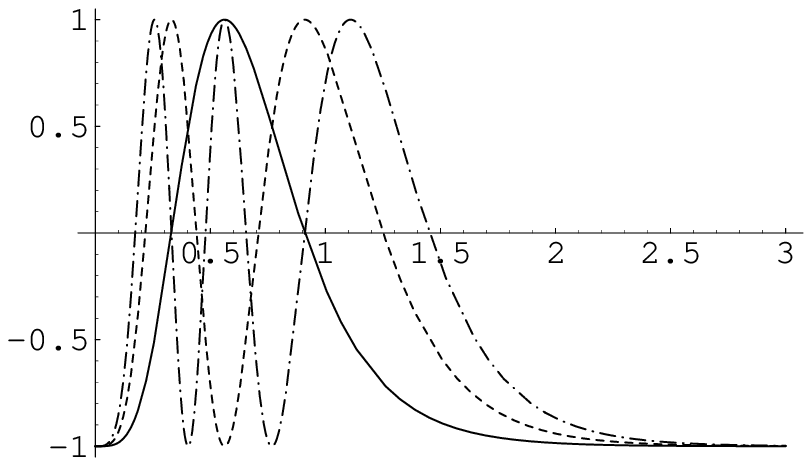}} {\bf
Fig. 3} \small{$ n^3 (\eta ) $ for $k=0,1,2$ - solid, dashed,
dot-dashed line respectively} \vspace{0.5cm}

\noindent Let us now calculate the Hopf index of the obtained
solutions. It can be done using the method presented in
\cite{aratyn}. We introduce new functions
\begin{equation}
\Phi_{\left( ^1 _2 \right)}=\left( \frac{f}{\sqrt{f^2+1}}\right)
\times \left( ^{\cos m\xi} _{\sin m \xi}  \right) \label{topPhi1}
\end{equation}
and
\begin{equation}
\Phi_{\left( ^3 _4 \right)}=\left( \frac{1}{\sqrt{f^2+1}}\right)
\times \left( ^{\cos n\phi} _{-\sin n \phi}  \right),
\label{topPhi3}
\end{equation}

\vspace{0.5cm} \centerline{\includegraphics{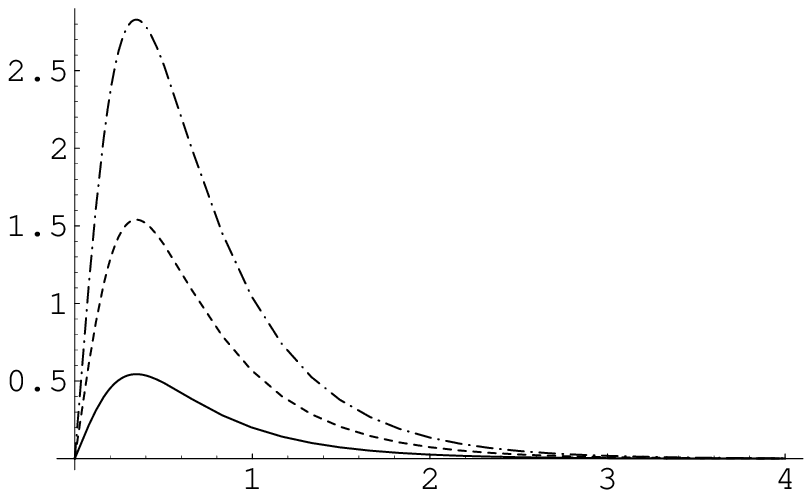}}
{\bf Fig. 4} \small{Energy density for $k=0,1,2$ - solid, dashed,
dot-dashed line respectively} \vspace{0.5cm}

\noindent which are connected with the unit vector field by the
relation $n_i = Z^{\dagger} \sigma_i Z$, where $\vec{\sigma}$ are
well-known Pauli matrices. Here
\begin{equation}
Z=\left(
\begin{array}{c}
Z_1 \\
Z_2
\end{array}
\right), \; \; \; Z^{\dagger} =( Z_1^*, Z_2^*)
\label{parametrHopf}
\end{equation}
and
\begin{equation}
Z_1=\Phi_1+i\Phi_2, \; \; \; Z_2=\Phi_3 +i\Phi_4.
\label{parametrZ}
\end{equation}
Because of the fact that Hopf index is defined by means of the
Abelian vector field and its derivatives
\begin{equation}
Q_H =\frac{1}{4\pi^2} \int d^3x \vec{A} \cdot \vec{B},
\label{indexHopf}
\end{equation}
where $\vec{B}=\vec{\nabla} \times \vec{A}$, we have to find
$\vec{A}$ as a function of the primary unit field. It can be done
and we get
\begin{equation}
A_i = \frac{i}{2} (Z^{\dagger} \partial_i Z -\partial_i
Z^{\dagger} Z). \label{ab-potential}
\end{equation}
Then, Hopf index can be evaluated and reads
\begin{equation}
Q_H=\frac{nm}{2} \sum_{i=0}^{l} \left[(\Phi^2_1+\Phi_2^2)^2 -
(\Phi_3^2+\Phi_4^2)^2 \right]_{\eta_i}^{\eta_{i+1}}, \label{hopf1}
\end{equation}
where, for odd $i$ $\eta_i$ is $i$-th singular point of the
function $f$ (\ref{sol3}) whereas for even value of $i$ $\eta_i$
is $i$-th zero of the function $f$. Of course, one can introduce
Hopf index for all soliton components of the solution. Namely,
\begin{equation}
Q_H^i=\frac{nm}{2} \left[(\Phi^2_1+\Phi_2^2)^2 -
(\Phi_3^2+\Phi_4^2)^2 \right]_{\eta_i}^{\eta{i+1}} = (-1)^i mn,
\label{hopf2}
\end{equation}
where $$ \eta_{2i+1}= \mbox{ar sinh} \left[ \frac{1+ \left(
\frac{|n|}{|m|} -1 \right) \frac{2i+1}{2l+1} }{ \sqrt{1- \left(
\left(1- \frac{|m|}{|n|} \right) \frac{2i+1}{2l+1}
+\frac{|m|}{|n|}  \right)^2 }} \right],$$ and
$$ \eta_{2i}= \mbox{ar sinh} \left[ \frac{1+ \left(
\frac{|n|}{|m|} -1 \right) \frac{2i}{2l+1} }{ \sqrt{1- \left(
\left(1- \frac{|m|}{|n|} \right) \frac{2i}{2l+1} +\frac{|m|}{|n|}
\right)^2 }} \right],$$
with $i=0,1...l$. \\
Finally we obtain that
\begin{equation}
Q_H=-mn. \label{hopf3}
\end{equation}
This result shows us that the obtained solution (\ref{sol3})
indeed consists of odd number of the toroidal solutions with
nontrivial topological charge. Each of the solitons corresponds to
the same absolute value of Hopf index, whereas the sign
oscillates. The total topological charge is constant and does not
depend on the number of oscillations. \\
Analogously, solution (\ref{sol_n3a}) is made of even numbers of
toroidal solitons with zero total topological charge.
\section{\bf{ Conclusions}}
In the present paper, a Lorentz invariant model based on the unit,
three component vector field has been investigated. This model
consists of two parts multiplying each other. Namely, the first
part, symmetric under the global $O(3)$ rotations and the second
which breaks this symmetry. The violating function (dielectric
function) has been chosen is the special form (\ref{sigma}). \\
It has been proved that such defined model possesses not only
standard toroidal solutions with arbitrary topological charge
known form recent work but also multi-soliton configurations.
Exact solutions, their energies and values of the Hopf index have
been obtained. In general, the solutions can be divided into two
classes with nonzero or zero total topological charge. The
solutions with nontrivial total Hopf index consist of odd numbers
nested toroidal solitons with a partial charge $\pm Q$ whereas
configurations with zero total Hopf index are build of even
numbers of such nested solitons. It has to be stressed that only
the most nested soliton i.e. located at $\eta=\infty$ is a
line-like object. Remaining hopfions are a little bit
pathological. They are two dimensional toruses with the
topological charge homogenously spread on their surface.
\\
Of course, because of the fact that all multi-soliton
configurations, with constant total Hopf index, have larger energy
than the standard one-soliton solution, we can expect that they
are unstable. Due to the Vakulenko-Kapitansky inequality one can
immediately see that all single solitons would attract each
others. Our knotted multi-soliton solution unties leading to the
stable one-soliton state. No\-ne\-the\-less, ob\-tai\-ned
mul\-ti-so\-li\-ton so\-lu\-tions (in par\-ti\-cu\-lar
soliton-antisoliton state) may give a chance to find a
breather-like state i.e. oscillating soliton with vanishing total
topological charge.
\\
It should be noticed that, as it was proved in \cite{my2},
investigated model is very unusual. It follows from the
observation that all dielectric functions $\sigma$ give the same
spectrum of the solitons i.e. their masses and topological charges
are identical and do not depend on the form of $\sigma$. The
dielectric function inflects only the shape of the hopfion. It is
true also in the case of here analyzed model, but only in the one
soliton sector that is $k=l=0$. Nested hopfions differ a lot from
standard Aratyn-Ferreira-Zimerman solutions. In addition to
different shapes they possess different total energy. Moreover,
since they consist of many solitons and antisolitons, the
topological contents of obtained configurations is also
dissimilar.
\\
However, because of the previously mentioned fact that one soliton
sector is identical for all $\sigma$ \cite{my2}, one can suppose
that such equivalence can be valid for multi-soliton sector as
well. If this conjecture were true then multi-soliton solutions
(and in particular sphalerons) should be observed in case of more
realistic dielectric functions \cite{my1}.
\\
To conclude, the existence of the multi-hopfions (sphaleron-like)
solutions, at least in the toy model, is very promising. In
particular, the problem of the breathers seems to be important in
the context of the Faddeev-Niemi model of glueballs where such
time-depending breathers could probably influence the spectrum of
the solutions i.e. expected spectrum of the glueball states. We
would like to address this problem in the forthcoming paper.
\vspace{5mm}

\noindent This work is par\-tial\-ly sup\-por\-ted by
Foun\-da\-tion for Pol\-ish Scie\-nce FNP and ESF "COS\-LAB"
programme.

\end{document}